# CEST MR-Fingerprinting: practical considerations and insights for acquisition schedule design and improved reconstruction

Or Perlman[1], Kai Herz[2,3], Moritz Zaiss[2], Ouri Cohen[4], Matthew S. Rosen[1,5*], Christian T. Farrar[1*]

[1]Athinoula A. Martinos Center for Biomedical Imaging, Department of Radiology, Massachusetts General Hospital and Harvard Medical School, Charlestown, MA, USA

[2]Magnetic Resonance Center, Max Planck Institute for Biological Cybernetics, Tübingen, Germany

[3]IMPRS for Cognitive and Systems Neuroscience, University of Tübingen, Tübingen, Germany

[4]Memorial Sloan Kettering Cancer Center, New York, NY, USA

[5]Department of Physics, Harvard University, Cambridge, MA, USA

**Grant sponsor:** National Institutes of Health; Grant number: R01CA203873

**Correspondence to:** Christian T. Farrar, Athinoula A. Martinos Center for Biomedical Imaging, Department of Radiology, Massachusetts General Hospital, 149 13th Street, Suite 2301, Charlestown, MA, 02129, USA. Email: cfarrar@nmr.mgh.harvard.edu

**Content:** Word count in text body: 4979; Figures: 8; Tables: 2; Supplementary figures: 1

**Submitted to Magnetic Resonance in Medicine**

---

*Christian T. Farrar and Matthew S. Rosen contributed equally to this work

**Purpose:** To understand the influence of various acquisition parameters on the ability of CEST MR-Fingerprinting (MRF) to discriminate different chemical exchange parameters and to provide tools for optimal acquisition schedule design and parameter map reconstruction.

**Methods:** Numerical simulations were conducted using a parallel-computing implementation of the Bloch-McConnell equations, examining the effect of TR, TE, flip-angle, water $T_1$ and $T_2$, saturation-pulse duration, power, and frequency on the discrimination ability of CEST-MRF. A modified Euclidean-distance matching metric was evaluated and compared to traditional dot-product matching. L-Arginine phantoms of various concentrations and pH were scanned at 4.7T and the results compared to numerical findings.

**Results:** Simulations for dot-product matching demonstrated that the optimal flip-angle and saturation times are 30° and 1100 ms, respectively. The optimal maximal saturation power was 3.4 $\mu$T for concentrated solutes with a slow exchange-rate, and 5.2 $\mu$T for dilute solutes with medium-to-fast exchange-rates. Using the Euclidean-distance matching metric, much lower maximum saturation powers were required (1.6 and 2.4 $\mu$T, respectively), with a slightly longer saturation time (1500 ms) and 90° flip-angle. For both matching metrics, the discrimination ability increased with the repetition time. The experimental results were in agreement with simulations, demonstrating that more than a 50% reduction in scan-time can be achieved by Euclidean-distance-based matching.

**Conclusion:** Optimization of the CEST-MRF acquisition schedule is critical for obtaining the best exchange parameter accuracy. The use of Euclidean-distance-based matching of signal trajectories simultaneously improved the discrimination ability and reduced the scan time and maximal saturation power required.

# 1. Introduction

Chemical Exchange Saturation Transfer (CEST) MRI is a molecular imaging technique that is capable of detecting milli-molar concentrations of exchangeable protons (1–4). The CEST contrast mechanism, when stemming from endogenous proteins and metabolites with exchangeable protons such as amine, amide, or hydroxyl groups, has provided clinical insights in a variety of disease pathologies. These include cancer (5–7), stroke (8, 9), mitochondrial disorders (10), disc and cartilage degeneration studies (11, 12), and neurodegenerative diseases (13–18). The same technique is even more sensitive when applied to exogenous materials (19) involving the use of paramagnetic lanthanides (19–21), liposomes (22), or iodine-containing substances (23).

Several challenges still prevent CEST-MRI from reaching its full potential to become a routine clinical imaging technique. First, the predominantly used CEST analysis method is the Magnetization Transfer Ratio asymmetry ($MTR_{asym}$ (24)), which is mixed with non-CEST contrast contributions and highly dependent on the acquisition protocol parameters:

$$MTR_{asym} = \frac{S(-\Delta\omega) - S(+\Delta\omega)}{S_0} \quad (1)$$

where $S(\pm\Delta\omega)$ is the signal measured with saturation at offset $\pm\Delta\omega$, and $S_0$ is the unsaturated signal. A recent review on the application of CEST to clinical scanners has shown that there is a large heterogeneity in the acquisition parameters used by different medical centers (25), making the comparison of findings difficult. Moreover, the $MTR_{asym}$ does not take under consideration the effect of the nuclear Overhauser enhancement (NOE) mediated aliphatic protons, which can be highly dominant in the brain, and is prone to contamination from the semi-solid magnetization transfer (MT) pool (26, 27) and water $T_1$ effects (28–31).

Ideally, for providing the most useful estimation of the metabolite of interest, the actual physical CEST properties – proton exchange-rate, and solute concentration should be mapped. In accordance, various efforts were previously taken to achieve quantitative CEST imaging (32) such as the quantitation of exchange using saturation power (QUESP) or time (QUEST) (33–35) and Omega plot (36) methods. These methods exploit the dependency of the CEST signal on the saturation power (or saturation time) and fit the $MTR_{asym}$ for a single offset as a function of the saturation parameter to estimate the labile proton volume fraction and exchange-rate. Although normalization of the $MTR_{asym}$ by the signal acquired at the negative offset frequency is intended to reduce the MT effect, it has been shown that the semisolid peak is actually asymmetric (37). Moreover, the QUESP type methods do not account for the NOE contribution. Alternatively, a multi-Lorentzian model can be fitted to the entire Z-spectrum, separating out the contribution of the CEST/MT/NOE pools (26, 38, 39). However, a single Z-spectrum-based Lorentzian fit provides only a semi-quantitative estimation of the pool features. To obtain the actual proton exchange-rates and concentrations a multi-saturation power acquisition is required, which can take from tens of minutes to more than an hour.

Magnetic resonance fingerprinting (MRF) is a new paradigm for quantitative imaging (40). Originally presented for quantitative mapping of water $T_1$, $T_2$ and $B_0$, this technique enables the fast and simultaneous mapping of several magnetic properties. It uses a pseudo-random acquisition schedule, which yields unique signal trajectories, capable of differentiating between various combinations of tissue properties. At the reconstruction step, the experimental data is compared to a Bloch-equation-based simulated dictionary, and the best match for each trajectory yields an estimated set of tissue properties (41). Recently, MRF was expanded and modified for CEST imaging (42–44) whereby the Bloch-McConnell equations are used to generate a reference dictionary, and a pseudo-random acquisition schedule with varied saturation power and/or times is used for obtaining signal trajectories and determine the exchange-rate and volume fraction of the solute of interest. In the realm of quantitative CEST imaging, CEST-MRF possess several important advantages: the acquisition time is much shorter than the alternatives (a few minutes); it takes under consideration the effect of various solute pools, without assuming any symmetry; and it can simultaneously output the fully quantitative properties of several-pools (42). Although preliminary results were promising, the incorporation of CEST-MRF in routine studies requires understanding the dependency of the discrimination ability on various pulse-sequence parameters. Unlike classic water-pool MRF, which relies on non-steady-state evolution of the magnetization, CEST requires considerable amplification of the labile proton signal and typically requires long saturation times leading to steady-state magnetization. Hence, it is essential to evaluate the limitations and technical considerations involving CEST-MRF acquisition schedules.

In the present study, the effect of various acquisition properties on the discrimination ability of CEST-MRF is systematically examined towards an optimized pulse-sequence. Moreover, a CEST-MRF specific Euclidian-distance-matching metric is suggested, and compared to the conventional dot-product metric, for improved parameter map reconstruction. We include numerical simulations in addition to in-vitro L-Arg phantom studies at 4.7T.

# 2. Methods

## 2.1 Simulated CEST scenarios

Three main representative CEST scenarios were numerically investigated, under 4.7T field conditions:

1. "Scenario A": Amide exchangeable solute proton with a slow exchange-rate but relatively high proton volume fraction, analogous to the endogenous amide protons observed in vivo. A three-pool simulation model was used that included the endogenous amide proton pool (chemical shift of 3.5 ppm), semi-solid proton pool (chemical shift of -2.5 ppm), and water proton pool.

2. "Scenario B": Amine exchangeable proton with a medium to fast chemical exchange-rate but relatively dilute proton volume fraction, analogous to applications such as imaging endogenous creatine (45), iodine-based pH probes (23), or CEST reporter genes (46). A two-pool case was simulated, with the solute chemical shift set at 3 ppm.

3. "Scenario C": To explore the general effect of water $T_1$ and $T_2$ changes on the optimized parameters, and to facilitate convenient validation using imaging phantoms, a third scenario was examined identical to scenario B but with longer relaxation times. Alterations in water $T_1$ and $T_2$ are expected in some clinical cases (e.g. edema), during the use of mixed iron-CEST agents (47), and at drastic pH changes involving exogenous materials (48). The simulated multi-pool properties for each of the three scenarios are detailed in Table 1.

## 2.2 Bloch-McConnell-based dictionary generation

Dictionaries of simulated signal intensity trajectories were generated using a Pade approximation (49) for the numerical solution of the Bloch-McConnell equations. A three pool system was simulated using three components for water, CEST, and semisolid MT as suggested previously (50, 51), forming a 7×7 matrix equation. For acceleration, the simulation was implemented in C++ using eigen (52) for linear algebra operations and openMP (53) for multi-threading. The source code was compiled with g++ 7.3 on an Ubuntu OS and is callable as a mex function in MATLAB (The MathWorks, Natick, MA).

## 2.3 Matching metric

The pattern-matching methodology, i.e., the assignment of the measured trajectory to a specific dictionary entry, determines the inherent discrimination of a given schedule. In this study, two matching metrics were used:

a. Vector dot product (DP) after 2-norm normalization (40):

$$DP(\mathbf{e},\mathbf{d}) = \frac{<\mathbf{e},\mathbf{d}>}{||\mathbf{e}||\cdot||\mathbf{d}||} \quad (2)$$

where $\mathbf{e}$ denotes an experimental signal trajectory and $\mathbf{d}$ denotes the dictionary entry vector.

b. Euclidean-distance (ED) with trajectory normalization by $M_0$ (the unsaturated reference signal) and the trajectory length ($N_t$):

$$ED(\mathbf{e},\mathbf{d}) = \frac{1}{\sqrt{N_t}}||\hat{\mathbf{Z}}_e - \hat{\mathbf{Z}}_d|| \quad (3)$$

where:

$$\hat{\mathbf{Z}}_e = \frac{\mathbf{e}}{M_{0e}}; \hat{\mathbf{Z}}_d = \frac{\mathbf{d}}{M_{0d}} \quad (4)$$

where $M_{0e}$ and $M_{0d}$ are the unsaturated reference signals for the experimental signal trajectory and the dictionary entry, respectively. Note that the normalization by $N_t$ does not have an effect on the optimization (for a given trajectory length) but is used to bound ED to the range [0, 1], as in DP.

## 2.4 Discrimination ability criteria

An ideal acquisition schedule will have a perfect match between the experimentally obtained trajectory and its ground-truth corresponding dictionary entry while having a poor match with any other entry. To quantify the discrimination ability we have used the following loss measures, each suitable for a specific matching metric:

1. Off-diagonal Frobenius norm dot-product loss (54) (for dot-product matching):

$$DP_{loss} = \frac{1}{N_D}||\mathbf{I} - \mathbf{D}^T\mathbf{D}||_f \quad (5)$$

where $\mathbf{D}$ is the dictionary dot-product matrix, consisting of the DP values for all $N_D$ combinations of dictionary entries, I is the identity matrix, and $||\ ||_f$ is the Frobenius norm. Intuitively, low DP$_{loss}$ values indicate that non-identical trajectories are close to orthogonal, hence the discrimination is optimized.

2. Off-diagonal Frobenius norm Euclidean-distance loss (for Euclidean-distance matching):

$$ED_{loss} = \frac{1}{N_D}||1 - \mathbf{I} - \mathbf{E}||_f \qquad (6)$$

where **E** is the dictionary Euclidean-distance matrix, containing ED values for all dictionary entries combination. The ED$_{loss}$ was designed to provide a qualitatively similar output to DP$_{loss}$, namely low values indicate better discrimination, while 1 indicates no discrimination.

3. Monte Carlo simulation of noise propagation. White Gaussian noise (25 dB) was added to the dictionary and the resulting trajectories matched to the original noiseless dictionary. The process was repeated 100 times, and the root-mean-squared-errors (RMSE) for the exchange-rate and proton volume fraction matching were calculated. This measure was used for an acquisition schedule truncation study whereby the number of schedule iterations was optimized as it has been recently found useful for that purpose (55). Moreover, DP$_{loss}$ and ED$_{loss}$ are not directly comparable and ED is biased when the number of iterations N$_t$ is varied.

## 2.5 Examining the dependence of the discrimination ability on the acquisition parameters

The influence of various acquisition parameters on the discrimination ability was numerically investigated. Since a relatively large parameter space affects the obtained signals, we focused the evaluations on two varied parameters at each step (while keeping all others fixed). The baseline acquisition schedule was set to the published version (42), which had a pseudo-random sequence of 30 saturation powers in the range of 0-6 $\mu$T, TR = 4s, TE = 21 ms, flip angle (FA) = 60°, saturation time (T$_{sat}$) = 3s, and a saturation offset frequency fixed to the solute offset frequency. The baseline saturation powers and the acquisition parameters examined are shown in Supporting Information Figure S1. Initially, the joint effect of varying the maximal saturation power and T$_{sat}$ was examined by rescaling the entire baseline schedule to have a maximum varying from 0.2 to 6 $\mu$T in 0.2 $\mu$T increments and T$_{sat}$ from 100 to 3900 ms in 100 ms increments. Next, the optimal B1$_{max}$ and T$_{sat}$ values were used, and the FA and TR were varied from 5° to 90° in 5° increments and from 100 ms higher than $T_{sat}$ to 8s in 100 ms increments, respectively. Finally, the optimal $B1_{max}$ and $T_{sat}$ values were again used with TR and FA fixed to their baseline values, but the saturation offset varied between 1 ppm lower than the solute offset to 1 ppm higher than the solute offset, in 0.1 ppm increments and the TE varied between 20 to 100 ms in 10 ms increments. For each combination of varied parameters, the $DP_{loss}$ and the $ED_{loss}$ were calculated, and the respective 3-D surface plot with its projected loss iso-contour lines was examined.

## 2.6 Optimization of the schedule length

To investigate the feasibility of reducing the number of schedule iterations and thus further shorten the acquisition time, dictionaries for the baseline schedule were re-created with $N_t$ varied from 1 to 30. This same schedule was used for both matching metrics to facilitate easy comparison. The $DP_{loss}$ was then calculated to predict the discrimination ability using the dot-product. We note that the equivalent calculation for $ED_{loss}$ is biased by $N_t$ and therefore cannot be used to optimize the schedule length. To nonetheless compare the predicted performance for both matching methods, a Monte Carlo simulation of noise propagation was performed.

## 2.7 Phantom study

The aim of the phantom study was to test the validity of the optimal acquisition parameters predicted by the loss measures, by experimentally performing the schedule length optimization study depicted above. A set of three L-arginine phantoms were used, similarly to (42), containing a total of 9 vials of 25-100 mM dissolved L-Arg in a buffer titrated to a range of 4-6 pH. The vials were surrounded by 2% agarose gel and imaged at room-temperature. Single-slice, single-shot CEST-MRF spin-echo EPI was acquired on a 4.7T MRI (Bruker, MA), with a 35-mm inner diameter birdcage volume coil (Bruker Biospin). The baseline acquisition schedule parameters (section 2.5) were used, with the addition of a preceding $M_0$ scan. For direct comparison between the Euclidean-distance and the dot-product metrics reconstructions from the same scan, the $M_0$-scan was followed by a single 15s repetition time.

As a reference ground-truth, quantitative estimation of the solute properties was performed using QUESP (33), employing an EPI schedule with TE = 21 ms, TR = 10s, FA = 90°, at saturation frequency offsets of ±3 ppm and 0-6 $\mu$T powers in 1 $\mu$T increments. $T_1$ maps were generated using variable repetition time images, acquired with TR = 200, 400, 800, 1500, 3000, and 5000 ms, TE = 7.5 ms, and FA = 90°. $T_2$ maps were generated from multi-echo spin-echo images, with a FA = 90°, TE = 9 ms, TR = 2000 ms, and 25 echoes between 9 and 225 ms. All imaging protocols had an identical geometry with the FOV set to 37×37 mm, and an isotropic pixel size of 1 mm.

## 2.8 Experimental data analysis

The CEST-MRF data was reconstructed into quantitative exchange-rate and concentration maps by pixel-wise matching the experimental trajectories to a dictionary comprised of the parameter combinations appearing in Table 1, "scenario C". The solute exchange-rate range was extended to 1400 Hz, to account for the high pH L-Arg vials (42). The dot-product and the Euclidean-distance metrics were both employed, with the normalization performed as described in section 2.3. $T_1$ and $T_2$ exponential fitting were performed using a custom-written program. To obtain ground-truth exchange-rate values, pixel-wise exchange-rate fitting of the QUESP data was performed with the known solute concentrations and measured water $T_1$ given as fixed inputs (34). For comparison, simultaneous fitting of the QUESP data for both the exchange-rate and concentration was also performed, by allowing both parameters to vary. Finally, the RMSE between the CEST-MRF exchange-rate and concentration maps and the respectively measured concentration and QUESP exchange-rates (estimated with input ground-truth concentrations) were calculated, using regions of interest (ROIs) of 36 mm$^3$. The mean±SD RMSE for all 3 phantoms was calculated for each schedule length case. Differences were evaluated by Student's t-test with p<0.05 considered as statistically significant. All calculations and fittings were performed using MATLAB.

## 3. Results

### 3.1 Dictionary simulation

A compiled parallel-computing implementation of the Bloch-McConnell equation simulations was used to generate the MRF dictionaries, comprised of the parameter combinations described in Table 1. The synthesis times was approximately 8 times faster than using the previously published sparse matrix implementation (42) on the same computer. For example, generation of the ∼670,000 entry dictionary described in (42) took 7.64 min, instead of 61 min on an Intel Xeon desktop computer equipped with four 2.27 GHz CPUs. It should be noted that the synthesis time could be further shortened by using a computer with more CPU cores.

## 3.2 The dependence of the discrimination ability on the acquisition parameters

The surface plots describing the discrimination ability for the dot-product metric are presented in Figure 1. The optimal saturation times were 1100-1200 ms for scenarios A and B and 2600 ms for scenario C, which simulated longer water $T_1$ and $T_2$. The optimal maximum saturation powers were 3.4, 5.2, and 6 $\mu$T respectively, for scenarios A, B, and C. In all three scenarios, the discrimination increased (loss decreased) with increasing flip-angle till approximately 20° and then plateaued with very slightly increasing loss for greater flip-angles. The discrimination ability continuously improved with increased repetition times in all cases. The echo time had no distinct effect on the loss, causing only slight variations. For the amide and semi-solid scenario A, the optimal saturation frequency offset was the same as the amide pool frequency (3.5 ppm), as expected. Interestingly, the optimal offset shifted 0.1 ppm from the simulated solute offset for scenario B, and 0.4 ppm for scenario C.

The surface plots describing the parameter discrimination ability for the Euclidean-distance metric are presented in Figure 2. The optimal saturation time was 1500 ms for scenario A, and 1600 ms for both scenario B, and C. Similar to the dot-product optimization, the optimal maximum saturation power increased from case A to B, and C, although the required powers were lower (1.6, 2.4, and 5.2 $\mu$T, respectively). The optimal TR was again 8000 ms for all scenarios, although a clear flip-angle dependency was evident here, yielding minimal ED$_{loss}$ for FA = 90°. The echo time had a minor influence on the loss, as can also be inferred from the straight and parallel loss iso-contours (Figure 2g-i). This may stem from the trajectories normalization and is in agreement with previous reports on the influence of TE on the CEST effect (56). A minimal echo time should, nonetheless, be chosen for optimal experiment SNR (TE = 21 ms was obtained for most cases in Figure 1,2g-i). The optimal saturation frequency was the same as the solute frequency for scenarios A and B, with a slight 0.1 ppm shift for scenario C.

The morphological differences between trajectories of various acquisition parameter combinations are shown in Figure 3 (for the dot-product metric), and Figure 4 (for the Euclidean-distance metric). Visually, the differences in trajectories for various CEST properties mostly manifested as amplitude scaling rather than distinctly different patterns. For both metrics, a pronounced deviation from the optimal set found was manifested as smaller amplitude differences between trajectories, accompanied by a reduced loss value. Although not optimal, the trajectories of the baseline acquisition schedule were relatively similar in amplitude (and in the resulting loss) to the best set of parameters, explaining the previously good results reported using this acquisition parameter set (42).

### 3.3 Optimizing the schedule length

The resulting DP$_{loss}$ values for different schedule lengths are shown in Figure 5a. A step-shaped improvement in the discrimination ability was demonstrated, with a leap in performance at $N_t$ = 11. The average RMSE values for matching the solute concentration are shown in Figure 5b. The dot-product related RMSE presented a similar step-like shape, with a similar leap at $N_t$ = 11. The Euclidean-distance RMSE were lower than that of the dot-product RMSE for most schedule lengths. Similar to the dot-product results, the Euclidean-distance RMSE predicted a discrimination ability improvement at $N_t$ = 11. However, the general convergence to the minimum RMSE was much faster, with less discrimination improvement after the 11th iteration (milder slope). For the solute exchange-rate ($K_{sw}$) (Figure 5c), the Euclidean-distance RMSE was again lower than the dot-product RMSE for short schedule lengths, but the errors converged to a similar or slightly higher value at the final iteration.

### 3.4 Phantom study

The measured solute concentrations and the QUESP exchange-rate images (generated with the known concentrations as input) for the 9 imaged vials are shown in Figure 6a. The dot-product matching of the CEST-MRF trajectories using 4 schedule iterations yielded poor results, with only a few vials matched correctly (Figure 6b). When 11 iteration-long trajectories were used, the results have improved, although significant errors are still visible. Using all 30 iterations, the errors are further reduced, yielding a more similar output to QUESP results. The corresponding results for the Euclidean-distance-based matching are shown in Figure 7. As can clearly be seen, using this metric the images converge to the QUESP results much faster, as most noticeable at $N_t$ = 11. The visual difference between the matching outcomes using 11 or 30 iterations is barely visible (as predicted by the numerical simulation in Figure 5). The quantitative analysis of the experimental phantom RMSE, compared to the reference QUESP images is shown in Figure 8. The RMSE for the Euclidean-distance matched images of solute concentration is significantly reduced at $N_t = 11$ compared to only 4 iterations, whereas using 30 iterations has not yielded significant improvement. The Euclidean-distance-based solute concentration RMSE are also significantly lower than the corresponding dot-product RMSE at $N_t = 11$. The dot-product-based solute concentration RMSE was not significantly reduced at $N_t = 11$ compared to $N_t = 4$ but was significantly reduced at $N_t = 30$. Although similar trends were obtained for the exchange-rate RMSE, no significant differences were observed. The quantitative values for all phantom vials are reported in Table 2.

## 4. Discussion

The optimization of CEST-MRF involves two competing mechanisms. On the one hand, traditional CEST imaging generally favors steady-state-like, high solute-signal conditions. On the other hand, classical MRF is typically characterized by low-SNR non-steady-state rapidly changing spin dynamics. The optimal parameters found (Figure 1-2), contain elements from both concepts. While the resulting saturation times (1100-1600 ms) were shorter than those typically used in continuous-wave pulse saturation CEST experiments (23), the optimal repetition time was at least three times longer than the water $T_1$. Nevertheless, to retain a short acquisition time with satisfying accuracy, a compromise in TR duration (TR = 4s) can be considered, with only a small sacrifice in discrimination ability. The compensation for speed-loss may come from the small number of schedule iterations required; 11-30 iterations for CEST-MRF. The inherent conflict between non-steady-state conditions and the requirement for sufficient CEST-SNR was also evident in the trajectories associated with the various sets of schedule parameters (Figure 3-4). While the patterns become more distinct for shorter saturation times, indicating the conditions are further away from steady-state, the signal amplitudes get smaller and the loss is actually increased, corresponding to decreased parameter discrimination (Figure 3c, f).

The optimal acquisition parameters found in this study are in general agreement with previously published CEST-MRF sequences. For example, the saturation power and time used for dot-product matching by Cohen et al. (42) ($T_{sat}$ = 3s, maximal B1 = 6 $\mu$T) are very close to the optimal parameters found here ($T_{sat}$ = 2.6s, $B_1$ = 6 $\mu$T) for a similar CEST scenario and are on the same loss iso-contour (Figure 1c). The optimal TE and FA are also very similar (FA = 60°, TE = 21 ms in (39); FA = 65°, TE = 21 ms in this study). Importantly, the optimization for the amide proton imaging 3-pool scenario, demonstrated that lower saturation power and times could have been used in previous brain in-vivo CEST-MRF efforts (42). Interestingly, in some cases (Figure 1h-i, Figure 2i) the optimal saturation frequency offset was slightly shifted with respect to the solute offset, suggesting that incorporating several saturation offsets in the schedule could be useful. The possible origin of these shifts is the fast exchange-rate involved, causing a broader CEST peak that is shifted closer to the water peak. The comparison with the results published in (43) is more difficult, as a different range of exchange-rates was imaged (0-600 Hz), no matching of the solute concentration was performed, and the T$_{sat}$ was varied during the acquisition. Although CEST-MRF can vary several acquisition parameters simultaneously at each iteration, we have chosen to focus this report on schedules that vary only the saturation power, due to the previously established efficiency of quantitative mapping with varied powers (34), and to simplify the multi-parameter dependencies involved. The tools used throughout this work (DP$_{loss}$, ED$_{loss}$, and the Monte-Carlo noise-based simulations) could be useful for future optimizations, as they are not dependent on the number of varied parameters at each iteration. Recently, a quantitative exchange-rate and concentration mapping method was suggested, which fits the signals of a steady-state CEST repeated experiment to the Bloch-McConnell equations using a nonlinear least-square technique (44). The authors reported that a CEST-MRF matching using the same parameters had inferior results. This can possibly be explained by the specific set of acquisition parameters used (T$_{sat}$ < 800 ms, saturation power < 1.2 $\mu$T) that clearly deviated from the optimal sets found here.

The dot-product matching metric is commonly used in MRF experiments (42, 43). However, several studies have recently used the Euclidean-distance metric (57, 58). Its utilization in this work, combined with the normalization by the M$_0$ signal, has demonstrated several important advantages. The optimal saturation powers for the Euclidean-distance metric were approximately two-times smaller than their equivalents for the dot-product metric, in both scenarios "A" and "B", and approximately 10% smaller for scenario "C" (Figures 1-2). This may reduce the specific absorption rate (SAR) level, an essential element for clinical translation. In both numerical simulations (Figure 5) and phantom studies (Figure 6-7), it was shown that the Euclidean-distance may reduce the matching errors, as well as reduce the schedule length (11 instead of 30 iterations) and hence scan time. We assume that the improved discrimination ability, demonstrated in both the numerical simulation (Figure 5) and in the phantom study (Figure 8), mostly at image acquisition number N$_t$ = 11, arises from the added information provided by the relatively low saturation power at this iteration (Supporting Information Figure S1b), which broadens the range of saturation powers used up until that iteration. In the phantom study con-ducted here, a single long TR (15s) was used following the added unsaturated reference scan, to allow convenient comparison with dot-product matching based on the same acquired images. However, a much shorter TR value can be used following this iteration, as the signal evolution is anyway simulated in the dictionary. The improved results gained by the Euclidean-distance metric can potentially be explained by examining the morphology of the MRF trajectories (Figures 3-4), showing that the discriminative information seems to be mostly manifested as signal intensity variations. As the Euclidean-distance metric is more sensitive to such information than to pattern variations (58) it seems to be more suitable for CEST-MRF than the dot-product metric. Moreover, the normalization by the unsaturated $M_0$ signal prevents the loss of some amplitude-related information, as caused by the 2-norm normalization of the entire acquired trajectory. To allow the flexibility of using both metrics, we suggest acquiring an $M_0$ scan at the beginning of the CEST-MRF schedule.

Interestingly, simultaneous fitting of QUESP data for the determination of exchangeable proton concentration and the exchange-rate yielded considerable errors in the solute concentration (Table 2). The total RMSE for the solute concentration was 16.4 mM for QUESP, compared to 8.03 mM for Euclidean-distance-based CEST-MRF with 30 acquisition iterations. This highlights the added value of fingerprinting as a multi-parameter matching method.

Another practical consideration demonstrated by the results obtained here is that the discrimination ability of CEST-MRF decreases with decreasing $T_2$ and increasing MT proton volume fraction. This is demonstrated in Figure 2 where the optimal discrimination is decreased (increased loss, min $ED_{loss}$ = 0.911) in scenario B with short water $T_2$ compared to scenario C (min $ED_{loss}$ = 0.824) with longer relaxation times. Similarly, the introduction of the semi-solid proton pool in scenario A leads to a further loss of exchange-rate discrimination (min $ED_{loss}$ = 0.950) compared to scenario B with no MT pool. The reduced discrimination observed for shorter $T_2$ and larger proton MT volume fraction can be overcome with higher SNR, so that small signal trajectory differences can still be distinguished, or larger ranges of chemical exchange-rate. This suggests that CEST-MRF will have better performance for exogenous CEST agents with fast chemical exchange-rates, compared to endogenous amide-proton imaging, and that longer $T_2$ relaxation times at lower field strengths may be advantageous.

Although the RMSE Monte-Carlo-based noise measure (Figure 5b-c) were generally able to predict the quantitative experimental results (Figure 8), some discrepancies were observed. These differences are mostly associated with the fact that the Monte Carlo simulations were performed for all dictionary entries throughout the entire range of parameters, whereas the experimental evaluation had a total of 9 vials, corresponding to 9 specific combinations of parameters. The noise level added to the Monte Carlo simulation (25 dB) was in good agreement with the actual noise measured from the 9 vials (23.8 ±1.93 dB). Since no image-averaging is performed in CEST-MRF, the noise level is slightly higher than typical CEST contrast (59). It should be noted that a Gaussian, rather than a Rician noise was used, due to the sufficiently high SNR levels (60, 61).

The usefulness of the DP$_{loss}$ metric was previously demonstrated in optimizing the varied schedule parameters (FA and TR) for water-pool MRF EPI sequences (54). However, Sommer et al. (55), have recently reported that dot-product-based metrics are not suitable for schedule truncation studies in classical water pool MRF, and demonstrated the efficiency of Monte-Carlo noise-based simulations for this task. Conversely, this study had yielded a generally good agreement between the DP$_{loss}$ and the Monte Carlo related measures (Figure 5a,b). A possible explanation may be the difference between the exact definitions of the loss. While a global maximum dot-product-based metric or a sum of dot-product correlations in only a small local area of the entire dictionary was used by Sommer et al. (55), a full consideration of the entire dictionary dot-product values was performed in DP$_{loss}$, as defined in this work. Moreover, while conventional T$_1$/T$_2$ MRF is prone to under-sampling noise (60), which are not accounted for in the dot-product-based loss measures, CEST-MRF is much less affected by such contaminations since a fully k-space sampled EPI sequence was employed with a relatively high SNR observed in the raw images.

Various efforts were previously taken to optimize the sequence of acquisition schedule parameters for T$_1$/T$_2$ water pool MRF (FA and TR) (54, 55, 62). The scope of this work was limited to a fundamental understanding of the limitations in CEST-MRF acquisition schedule parameters range, and to reducing the number of schedule iterations and thus used a fixed schedule with the range of the saturation power scaled. Nevertheless, the sequence of saturation powers used in a CEST-MRF experiment could be further optimized using the above-mentioned published methods combined with the fast dictionary generation software and the discrimination metrics presented here.

# 5. Conclusion

CEST-MRF holds unique challenges for the optimization of the image acquisition parameters stemming from the long saturation times required to generate significant CEST contrast. Here, we found optimal acquisition parameters that represent a compromise between generating large amplitude differences in the signal trajectory and non-steady-state conditions with unique trajectory patterns. The Euclidean-distance-based matching of signal trajectories may simultaneously improve the discrimination ability and reduce the scan time. For obtaining the most accurate CEST-MRF exchange parameter maps, it is critical to optimize the acquisition parameters for the specific application using numerical simulations of the parameter discrimination.

# References

[1] Van Zijl Peter CM, Yadav Nirbhay N. Chemical exchange saturation transfer (CEST): what is in a name and what isn't? *Magnetic resonance in medicine.* 2011;65:927–948.

[2] Ward KM, Balaban RS. Determination of pH using water protons and chemical exchange dependent saturation transfer (CEST) *Magnetic Resonance in Medicine: An Official Journal of the International Society for Magnetic Resonance in Medicine.* 2000;44:799–802.

[3] Zaiss Moritz, Bachert Peter. Chemical exchange saturation transfer (CEST) and MR Z-spectroscopy in vivo: a review of theoretical approaches and methods *Physics in Medicine & Biology.* 2013;58:R221.

[4] Liu Guanshu, Song Xiaolei, Chan Kannie WY, McMahon Michael T. Nuts and bolts of chemical exchange saturation transfer MRI *NMR in Biomedicine.* 2013;26:810–828.

[5] Zhou Jinyuan, Tryggestad Erik, Wen Zhibo, et al. Differentiation between glioma and radiation necrosis using molecular magnetic resonance imaging of endogenous proteins and peptides *Nature medicine.* 2011;17:130.

[6] Jones Craig K, Schlosser Michael J, Van Zijl Peter CM, Pomper Martin G, Golay Xavier, Zhou Jinyuan. Amide proton transfer imaging of human brain tumors at 3T *Magnetic Resonance in Medicine: An Official Journal of the International Society for Magnetic Resonance in Medicine.* 2006;56:585–592.

[7] Zhou Jinyuan, Blakeley Jaishri O, Hua Jun, et al. Practical data acquisition method for human brain tumor amide proton transfer (APT) imaging *Magnetic Resonance in Medicine: An Official Journal of the International Society for Magnetic Resonance in Medicine.* 2008;60:842–849.

[8] Sun Phillip Zhe, Cheung Jerry S, Wang Enfeng, Lo Eng H. Association between pH-weighted endogenous amide proton chemical exchange saturation transfer MRI and tissue lactic acidosis during acute ischemic stroke *Journal of Cerebral Blood Flow & Metabolism.* 2011;31:1743–1750.

[9] Zhou Jinyuan, Zijl Peter CM. Defining an acidosis-based ischemic penumbra from pH-weighted MRI *Translational stroke research.* 2012;3:76–83.

[10] DeBrosse Catherine, Nanga Ravi Prakash Reddy, Wilson Neil, et al. Muscle oxidative phosphorylation quantitation using creatine chemical exchange saturation transfer (CrCEST) MRI in mitochondrial disorders *JCI insight.* 2016;1.

[11] Singh Anup, Haris Mohammad, Cai Kejia, et al. Chemical exchange saturation transfer magnetic resonance imaging of human knee cartilage at 3 T and 7 T *Magnetic resonance in medicine.* 2012;68:588–594.

[12] Müller-Lutz Anja, Schleich Christoph, Pentang Gael, et al. Age-dependency of glycosaminoglycan content in lumbar discs: A 3t gagcEST study *Journal of Magnetic Resonance Imaging.* 2015;42:1517–1523.

[13] Bagga Puneet, Pickup Stephen, Crescenzi Rachelle, et al. In vivo GluCEST MRI: Reproducibility, background contribution and source of glutamate changes in the MPTP model of Parkinson's disease *Scientific reports.* 2018;8:2883.

[14] Cai Kejia, Haris Mohammad, Singh Anup, et al. Magnetic resonance imaging of glutamate *Nature medicine.* 2012;18:302.

[15] Davis Kathryn Adamiak, Nanga Ravi Prakash Reddy, Das Sandhitsu, et al. Glutamate imaging (GluCEST) lateralizes epileptic foci in nonlesional temporal lobe epilepsy *Science translational medicine.* 2015;7:309ra161–309ra161.

[16] Pépin Jérémy, Francelle Laetitia, Sauvage Maria-Angeles, et al. In vivo imaging of brain glutamate defects in a knock-in mouse model of Huntington's disease *Neuroimage.* 2016;139:53–64.

[17] Crescenzi Rachelle, DeBrosse Catherine, Nanga Ravi Prakash Reddy, et al. In vivo measurement of glutamate loss is associated with synapse loss in a mouse model of tauopathy *Neuroimage.* 2014;101:185–192.

[18] Crescenzi Rachelle, DeBrosse Catherine, Nanga Ravi PR, et al. Longitudinal imaging reveals subhippocampal dynamics in glutamate levels associated with histopathologic events in a mouse model of tauopathy and healthy mice *Hippocampus.* 2017;27:285–302.

[19] Zhang Shanrong, Merritt Matthew, Woessner Donald E, Lenkinski Robert E, Sherry A Dean. PARACEST agents: modulating MRI contrast via water proton exchange *Accounts of chemical research.* 2003;36:783–790.

[20] Aime Silvio, Barge Alessandro, Delli Castelli Daniela, et al. Paramagnetic lanthanide (III) complexes as pH-sensitive chemical exchange saturation transfer (CEST) contrast agents for MRI applications *Magnetic Resonance in Medicine: An Official Journal of the International Society for Magnetic Resonance in Medicine.* 2002;47:639–648.

[21] Wu Yunkou, Zhang Shanrong, Soesbe Todd C, et al. pH imaging of mouse kidneys in vivo using a frequency-dependent paraCEST agent *Magnetic resonance in medicine.* 2016;75:2432–2441.

[22] Terreno Enzo, Cabella Claudia, Carrera Carla, et al. From spherical to osmotically shrunken paramagnetic liposomes: an improved generation of LIPOCEST MRI agents with highly shifted water protons *Angewandte Chemie.* 2007;119:984–986.

[23] Longo Dario L, Sun Phillip Zhe, Consolino Lorena, Michelotti Filippo C, Uggeri Fulvio, Aime Silvio. A general MRI-CEST ratiometric approach for pH imaging: demonstration of in vivo pH mapping with iobitridol *Journal of the American Chemical Society.* 2014;136:14333–14336.

[24] Vinogradov Elena, Sherry A Dean, Lenkinski Robert E. CEST: from basic principles to applications, challenges and opportunities *Journal of magnetic resonance.* 2013;229:155–172.

[25] Jones Kyle M, Pollard Alyssa C, Pagel Mark D. Clinical applications of chemical exchange saturation transfer (CEST) MRI *Journal of Magnetic Resonance Imaging.* 2018;47:11–27.

[26] Zaiss Moritz, Windschuh Johannes, Paech Daniel, et al. Relaxation-compensated CEST-MRI of the human brain at 7 T: unbiased insight into NOE and amide signal changes in human glioblastoma *Neuroimage.* 2015;112:180–188.

[27] Zijl Peter CM, Lam Wilfred W, Xu Jiadi, Knutsson Linda, Stanisz Greg J. Magnetization Transfer Contrast and Chemical Exchange Saturation Transfer MRI. Features and analysis of the field-dependent saturation spectrum *Neuroimage.* 2018;168:222–241.

[28] Xu Junzhong, Zaiss Moritz, Zu Zhongliang, et al. On the origins of chemical exchange saturation transfer (CEST) contrast in tumors at 9.4 T *NMR in biomedicine.* 2014;27:406–416.

[29] Sun Phillip Zhe, Murata Yoshihiro, Lu Jie, Wang Xiaoying, Lo Eng H, Sorensen A Gregory. Relaxation-compensated fast multislice amide proton transfer (APT) imaging of acute ischemic stroke *Magnetic Resonance in Medicine: An Official Journal of the International Society for Magnetic Resonance in Medicine.* 2008;59:1175–1182.

[30] Khlebnikov Vitaliy, Polders Daniel, Hendrikse Jeroen, et al. Amide proton transfer (APT) imaging of brain tumors at 7 T: the role of tissue water T1-relaxation properties *Magnetic resonance in medicine.* 2017;77:1525–1532.

[31] Zu Zhongliang. Towards the complex dependence of MTRasym on T1w in amide proton transfer (APT) imaging *NMR in Biomedicine.* 2018;31:e3934.

[32] Ji Yang, Zhou Iris Yuwen, Qiu Bensheng, Sun Phillip Zhe. Progress toward quantitative in vivo chemical exchange saturation transfer (CEST) MRI *Israel Journal of Chemistry.* 2017;57:809–824.

[33] McMahon Michael T, Gilad Assaf A, Zhou Jinyuan, Sun Phillip Z, Bulte Jeff WM, Zijl Peter CM. Quantifying exchange rates in chemical exchange saturation transfer agents using the saturation time and saturation power dependencies of the magnetization transfer effect on the magnetic resonance imaging signal (QUEST and QUESP): pH calibration for poly-L-lysine and a starburst dendrimer *Magnetic Resonance in Medicine: An Official Journal of the International Society for Magnetic Resonance in Medicine.* 2006;55:836–847.

[34] Zaiss Moritz, Angelovski Goran, Demetriou Eleni, McMahon Michael T, Golay Xavier, Scheffler Klaus. QUESP and QUEST revisited–fast and accurate quantitative CEST experiments *Magnetic resonance in medicine.* 2018;79:1708–1721.

[35] Demetriou Eleni, Tachrount Mohamed, Zaiss Moritz, Shmueli Karin, Golay Xavier. PRO-QUEST: a rapid assessment method based on progressive saturation for quantifying exchange rates using saturation times in CEST *Magnetic resonance in medicine.* 2018;80:1638–1654.

[36] Wu Renhua, Xiao Gang, Zhou Iris Yuwen, Ran Chongzhao, Sun Phillip Zhe. Quantitative chemical exchange saturation transfer (qCEST) MRI–omega plot analysis of RF-spillover-corrected inverse CEST ratio asymmetry for simultaneous determination of labile proton ratio and exchange rate *NMR in biomedicine.* 2015;28:376–383.

[37] Hua Jun, Jones Craig K, Blakeley Jaishri, Smith Seth A, Van Zijl Peter CM, Zhou Jinyuan. Quantitative description of the asymmetry in magnetization transfer effects around the water resonance in the human brain *Magnetic Resonance in Medicine: An Official Journal of the International Society for Magnetic Resonance in Medicine.* 2007;58:786–793.

[38] Zaiss Moritz, Schmitt Benjamin, Bachert Peter. Quantitative separation of CEST effect from magnetization transfer and spillover effects by Lorentzian-line-fit analysis of z-spectra *Journal of magnetic resonance.* 2011;211:149–155.

[39] Zhou Iris Yuwen, Wang Enfeng, Cheung Jerry S, Zhang Xiaoan, Fulci Giulia, Sun Phillip Zhe. Quantitative chemical exchange saturation transfer (CEST) MRI of glioma using Image Downsampling Expedited Adaptive Least-squares (IDEAL) fitting *Scientific reports.* 2017;7:84.

[40] Ma Dan, Gulani Vikas, Seiberlich Nicole, et al. Magnetic resonance fingerprinting *Nature.* 2013;495:187.

[41] Bipin Mehta Bhairav, Coppo Simone, Frances McGivney Debra, et al. Magnetic resonance fingerprinting: a technical review *Magnetic resonance in medicine.* 2019;81:25–46.

[42] Cohen Ouri, Huang Shuning, McMahon Michael T, Rosen Matthew S, Farrar Christian T. Rapid and quantitative chemical exchange saturation transfer (CEST) imaging with magnetic resonance fingerprinting (MRF) *Magnetic resonance in medicine.* 2018;80:2449–2463.

[43] Zhou Zhengwei, Han Pei, Zhou Bill, et al. Chemical exchange saturation transfer fingerprinting for exchange rate quantification *Magnetic resonance in medicine.* 2018.

[44] Heo Hye-Young, Han Zheng, Jiang Shanshan, Schär Michael, Zijl Peter CM, Zhou Jinyuan. Quantifying amide proton exchange rate and concentration in chemical exchange saturation transfer imaging of the human brain *NeuroImage.* 2019.

[45] Haris Mohammad, Nanga Ravi Prakash Reddy, Singh Anup, et al. Exchange rates of creatine kinase metabolites: feasibility of imaging creatine by chemical exchange saturation transfer MRI *NMR in Biomedicine.* 2012;25:1305–1309.

[46] Farrar Christian T, Buhrman Jason S, Liu Guanshu, et al. Establishing the lysine-rich protein CEST reporter gene as a CEST MR imaging detector for oncolytic virotherapy *Radiology.* 2015;275:746–754.

[47] Gilad Assaf A, Laarhoven Hanneke WM, McMahon Michael T, et al. Feasibility of concurrent dual contrast enhancement using CEST contrast agents and superparamagnetic iron oxide particles *Magnetic Resonance in Medicine: An Official Journal of the International Society for Magnetic Resonance in Medicine.* 2009;61:970–974.

[48] Sun Phillip Zhe, Longo Dario Livio, Hu Wei, Xiao Gang, Wu Renhua. Quantification of iopamidol multi-site chemical exchange properties for ratiometric chemical exchange saturation transfer (CEST) imaging of pH *Physics in Medicine & Biology.* 2014;59:4493.

[49] Golub Gene H, Van Loan Charles F. *Matrix computations*;3. JHU Press 2012.

[50] Morrison Clare, Mark Henkelman R. A model for magnetization transfer in tissues *Magnetic resonance in medicine.* 1995;33:475–482.

[51] Zaiss Moritz, Zu Zhongliang, Xu Junzhong, et al. A combined analytical solution for chemical exchange saturation transfer and semi-solid magnetization transfer *NMR in Biomedicine.* 2015;28:217–230.

[52] Guennebaud Gaël, Jacob Benoît, others . Eigen v3 http://eigen.tuxfamily.org 2010.

[53] Dagum Leonardo, Menon Ramesh. OpenMP: An industry-standard API for shared-memory programming *Computing in Science & Engineering.* 1998:46–55.

[54] Cohen Ouri, Rosen Matthew S. Algorithm comparison for schedule optimization in MR fingerprinting *Magnetic resonance imaging.* 2017;41:15–21.

[55] Sommer K, Amthor T, Doneva M, Koken P, Meineke J, Börnert P. Towards predicting the encoding capability of MR fingerprinting sequences *Magnetic resonance imaging.* 2017;41:7–14.

[56] Sun Phillip Zhe, Wang Yu, Lu Jie. Sensitivity-enhanced chemical exchange saturation transfer (CEST) MRI with least squares optimization of Carr Purcell Meiboom Gill multi-echo echo planar imaging *Contrast media & molecular imaging.* 2014;9:177–181.

[57] Cline Christopher C, Chen Xiao, Mailhe Boris, et al. AIR-MRF: Accelerated iterative reconstruction for magnetic resonance fingerprinting *Magnetic resonance imaging.* 2017;41:29–40.

[58] Ma Heather T, Ye Chenfei, Wu Jun, et al. A preliminary study of DTI Fingerprinting on stroke analysis in *Engineering in Medicine and Biology Society (EMBC), 2014 36th Annual International Conference of the IEEE*:2380–2383IEEE 2014.

[59] Zhou Jinyuan, Heo Hye-Young, Knutsson Linda, Zijl Peter CM, Jiang Shanshan. APT-weighted MRI: Techniques, current neuro applications, and challenging issues *Journal of Magnetic Resonance Imaging.* 2019.

[60] Kara Danielle, Fan Mingdong, Hamilton Jesse, Griswold Mark, Seiberlich Nicole, Brown Robert. Parameter map error due to normal noise and aliasing artifacts in MR fingerprinting *Magnetic resonance in medicine.* 2019.

[61] Gudbjartsson Hákon, Patz Samuel. The Rician distribution of noisy MRI data *Magnetic resonance in medicine.* 1995;34:910–914.

[62] Zhao Bo, Haldar Justin P, Liao Congyu, et al. Optimal experiment design for magnetic resonance fingerprinting: Cramer-Rao bound meets spin dynamics *IEEE transactions on medical imaging.* 2018.

# Tables

Table 1: Relaxation time and chemical exchange parameters for the examined CEST application scenarios

| Scenario | A | B | C |
| --- | --- | --- | --- |
| Description | Three-pool amide proton and semisolid MT | Diluted solutes in the medium to fast exchange-rate regime (two-pool) | Diluted solutes in the medium to fast exchange-rate regime (two-pool) + long water relaxation times |
| Water $T_1$ (ms) | 1450 | 1450 | 2450 |
| Water $T_2$ (ms) | 50 | 50 | 600 |
| Solute $T_1$ (ms) | 1450 | 1450 | 2450 |
| Solute $T_2$ (ms) | 1 | 40 | 40 |
| $K_{sw}$ (Hz) | 5:5:150 | 5:5:1000 | 5:5:1000 |
| Solute concentration (mM) | 100:50:1000 | 10:5:120 | 10:5:120 |
| Exchangeable protons per solute | 1 | 3 | 3 |
| Off-set frequency (ppm) | 3.5 | 3 | 3 |
| Semi-solid $T_1$ (ms) | 1450 | — | — |
| Semi-solid $T_2$ (ms) | 0.04 | — | — |
| Semi-solid concentration (M) | 13.2 | — | — |
| $K_{ssw}$ (Hz) | 30 | — | — |

Table 2: Comparison of the CEST-MRF dot-product or Euclidean-distance matched L-Arg concentrations and amine proton chemical exchange-rates ($K_{ex}$) with QUESP measured values for the various phantom vials.

| Ground-truth concentration | QUESP with ground-truth concentration as input | Dot-product with $N_t = 4$ | | Dot-product with $N_t = 11$ | | Dot-product with $N_t = 30$ | |
|---|---|---|---|---|---|---|---|
| [L-arg] (mM) | $K_{ex}$ (Hz) | [L-arg] (mM) | $K_{ex}$ (Hz) | [L-arg] (mM) | $K_{ex}$ (Hz) | [L-arg] (mM) | $K_{ex}$ (Hz) |
| 50 | 104.6±8.4 | 100.6±13.4 | 82.4±2.5 | 120.0±0.0 | 70.8±5.8 | 56.3±4.5 | 105.1±17.8 |
| 50 | 138.7±13.7 | 109.6±11.0 | 84.0±2.8 | 118.3±6.9 | 92.1±8.8 | 61.7±4.0 | 136.8±15.9 |
| 50 | 240.7±17.3 | 93.2±34.9 | 106.4±47.3 | 79.2±29.4 | 218.5±86.0 | 59.0±8.5 | 248.2±62.0 |
| 100 | 268.1±30.3 | 80.7±28.6 | 305.6±106.2 | 89.6±13.3 | 365.3±62.3 | 88.9±11.8 | 369.6±62.0 |
| 25 | 206.7±29.3 | 74.8±15.3 | 83.9±2.1 | 109.3±16.3 | 79.0±13.7 | 39.3±3.6 | 148.9±19.9 |
| 50 | 262.3±24.5 | 93.8±25.9 | 87.2±3.9 | 82.5±24.4 | 202.8±62.6 | 64.6±8.2 | 235.3±46.0 |
| 50 | 569.3±47.4 | 47.5±11.7 | 540.1±91.6 | 60.1±2.2 | 524.4±38.0 | 60.1±2.2 | 526.7±37.6 |
| 50 | 255.2±26.3 | 93.2±31.8 | 124.0±124.2 | 71.4±20.4 | 241.7±80.3 | 59.9±7.7 | 267.6±67.4 |
| 50 | 887.4±144.9 | 52.9±25.4 | 1075.0±167.9 | 53.6±2.8 | 982.5±93.4 | 52.9±2.5 | 990.0±91.9 |
| **QUESP simultaneous estimation of [L-arg] and $K_{ex}$** | | **Euclidean-distance: $N_t = 4$** | | **Euclidean-distance: $N_t = 11$** | | **Euclidean-distance: $N_t = 30$** | |
| [L-arg] (mM) | $K_{ex}$ (Hz) | [L-arg] (mM) | $K_{ex}$ (Hz) | [L-arg] (mM) | $K_{ex}$ (Hz) | [L-arg] (mM) | $K_{ex}$ (Hz) |
| 32.0±3.5 | 171.7±19.4 | 56.4±9.8 | 82.8±7.2 | 44.7±4.9 | 114.7±17.1 | 44.0±6.0 | 120.6±21.0 |
| 31.6±3.1 | 233.6±20.0 | 79.7±16.0 | 83.5±9.5 | 43.9±5.6 | 159.3±18.1 | 42.4±6.4 | 171.1±23.5 |
| 34.9±2.7 | 374.9±33.4 | 97.8±20.4 | 130.3±50.3 | 47.2±5.1 | 273.9±63.0 | 46.8±4.7 | 282.4±61.1 |
| 67.7±10.3 | 446.8±49.1 | 98.9±17.8 | 311.5±76.3 | 81.7±8.0 | 381.9±55.8 | 81.1±7.7 | 390.1±54.8 |
| 13.1±1.9 | 473.3±36.7 | 43.2±8.1 | 82.5±10.2 | 20.8±2.5 | 186.0±25.1 | 20.0±2.4 | 205.3±26.3 |
| 37.1±4.5 | 383.4±51.0 | 108.5±8.8 | 120.6±16.6 | 56.9±6.4 | 246.5±47.1 | 56.5±6.7 | 256.5±47.0 |
| 48.6±3.1 | 597.1±56.0 | 58.3±6.1 | 509.3±78.5 | 55.3±1.2 | 546.3±37.4 | 55.6±1.6 | 554.0±43.4 |
| 37.2±4.1 | 367.4±29.2 | 100.7±21.5 | 140.8±88.2 | 46.0±4.6 | 296.7±57.1 | 45.6±4.6 | 307.9±59.0 |
| 45.8±3.4 | 1076.5±83.8 | 50.4±3.2 | 1022.4±143.8 | 50.3±2.9 | 1020.3±96.1 | 51.1±2.7 | 1003.8±114.4 |

# Figures

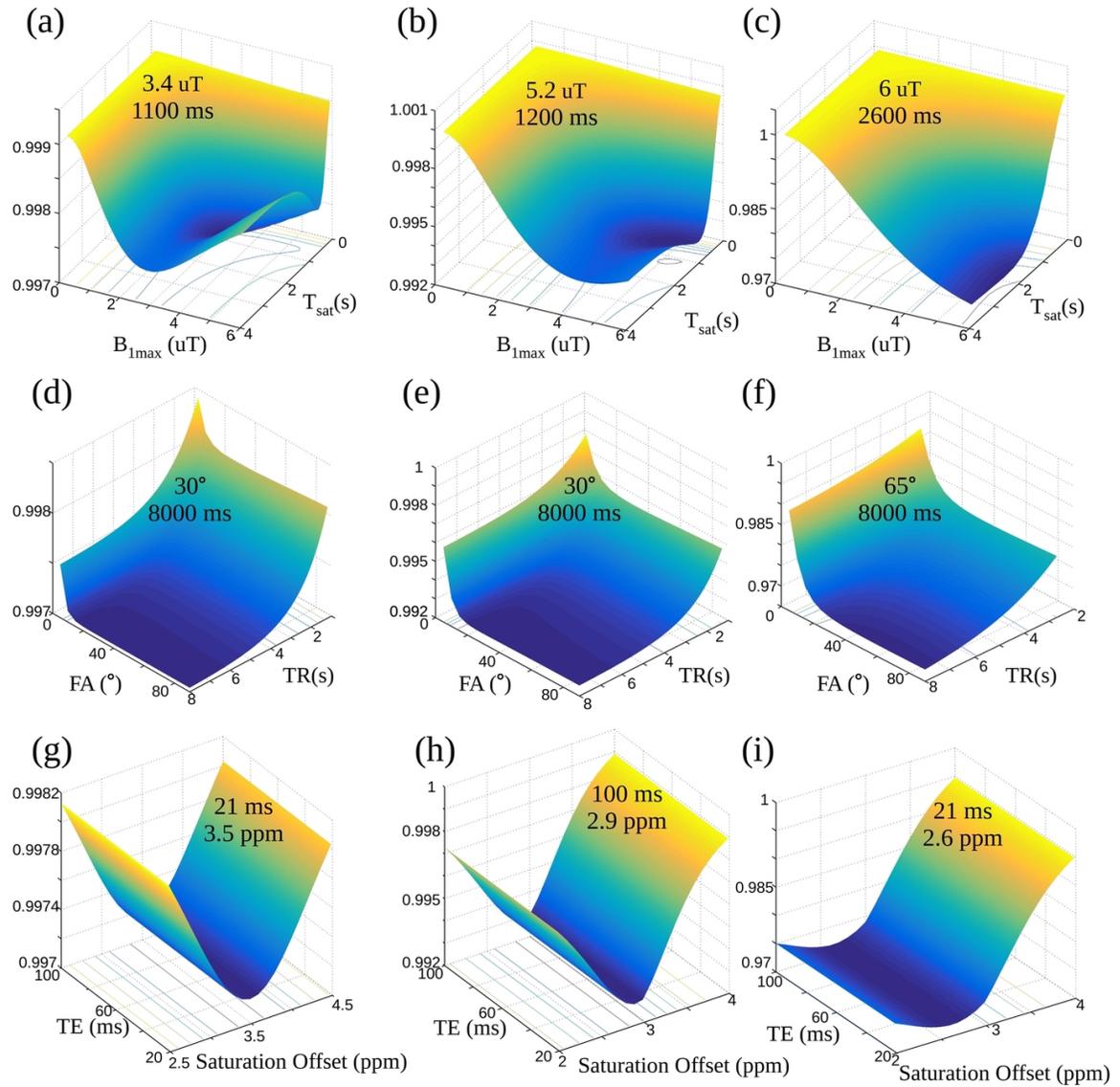

Figure 1: Dependence of the dot-product loss on the acquisition parameters. The surface plots with projected loss iso-contours describe the effect of the maximal saturation power and the saturation time ($T_{sat}$) (a-c), the flip angle (FA) and TR (d-f), and the TE and saturation frequency offset (g-i), on the $DP_{loss}$, for scenarios A (left column), B (center column), and C (right column). In all images, the z-axis represents the $DP_{loss}$, which is also color-coded from blue to yellow. The optimal combination for each examined parameter pair is given in the surface plot.

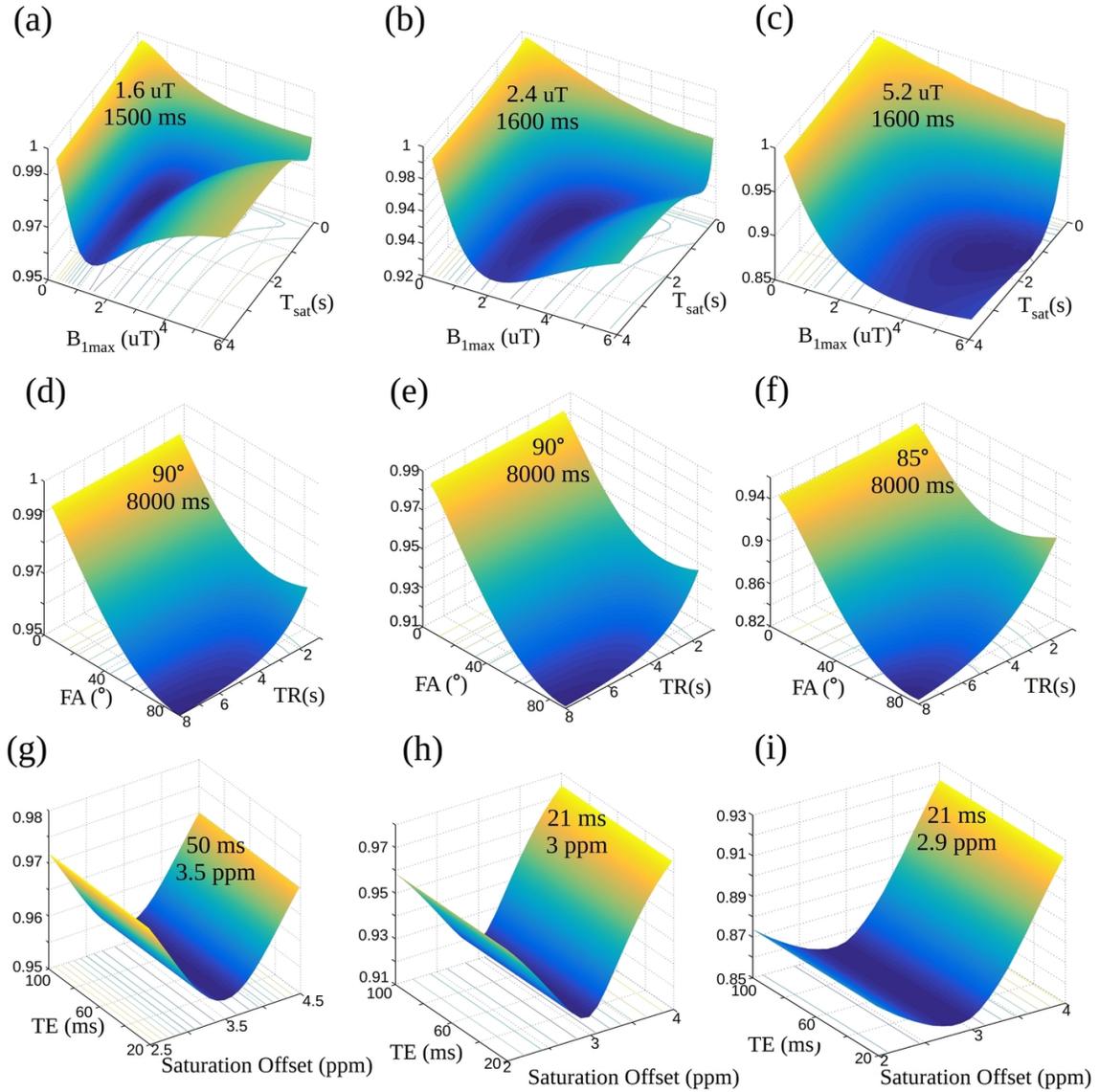

Figure 2: Dependence of the Euclidean-distance loss on the acquisition parameters. The surface plots with projected loss iso-contours describe the effect of the maximal saturation power and the saturation time ($T_{sat}$) (a-c), the FA and TR (d-f), and the TE and saturation frequency offset (g-i), on the $ED_{loss}$, for scenarios A (left column), B (center column), and C (right column). In all images, the z-axis represents the $ED_{loss}$, which is also color-coded from blue to yellow. The optimal combination for each examined parameter pair is given in the surface plot.

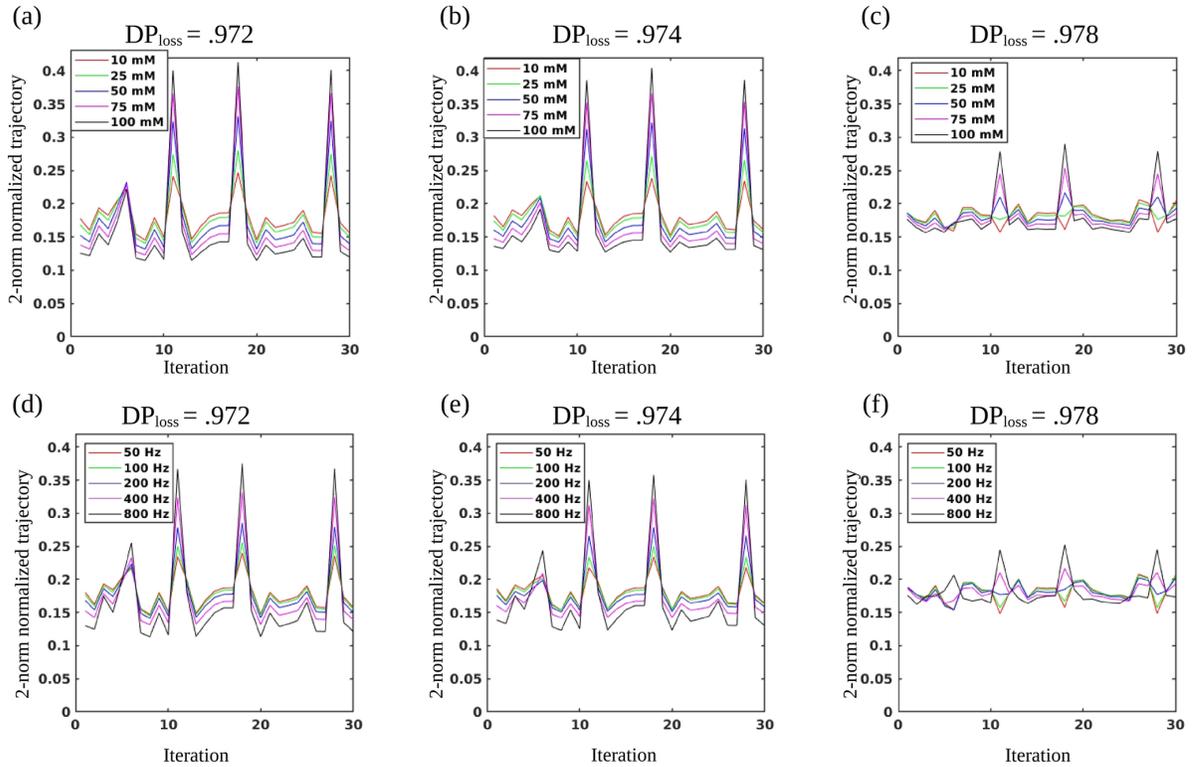

Figure 3: Exemplary trajectories for different combinations of acquisition schedule parameters normalized by the 2-norm (for dot-product matching). In (a-c), the exchange-rate was fixed at 400 Hz, and the solute concentration varied from 10 to 100 mM. In (d-f), the solute concentration was fixed on 50 mM, and the exchange-rate varied from 50 to 800 Hz. The left column represents a close to optimal acquisition parameters combination (TR was set to 4s instead of 8s for speed considerations), with saturation time ($T_{sat}$) = 2600 ms, FA = 60°, saturation offset = 2.6 ppm, TE = 21 ms, maximum saturation power = 6 $\mu$T. The center column represents the baseline acquisition schedule ($T_{sat}$ = 3000 ms, saturation offset = 3 ppm, see section 2.5), and the right column represents the same schedule, but with shorter saturation time and excitation flip angle (FA = 15°, and $T_{sat}$ = 1500 ms). The same CEST properties of scenario 'C' were used for all trajectories.

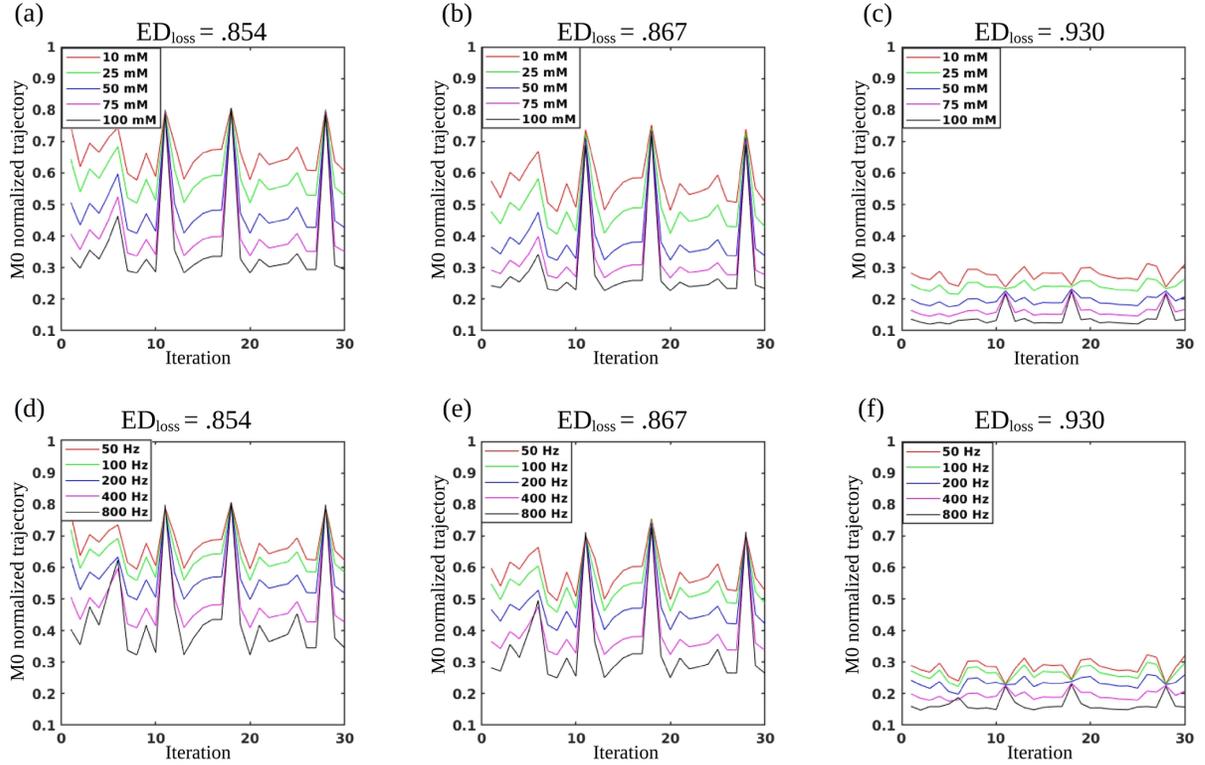

Figure 4: Exemplary trajectories for different combinations of acquisition schedule parameters normalized by $M_0$ (for Euclidean-distance matching). In (a-c), the exchange-rate was fixed at 400 Hz, and the solute concentration varied from 10 to 100 mM. In (d-f), the solute concentration was fixed on 50 mM, and the exchange-rate varied from 50 to 800 Hz. The left column represents a close to optimal acquisition parameters combination (TR was set to 4s instead of 8s for speed considerations), with saturation time ($T_{sat}$) = 1600 ms, FA = 90°, saturation offset = 2.9 ppm, TE = 21 ms, maximum saturation power = 5.2 $\mu$T. The center column represents the baseline acquisition schedule ($T_{sat}$ = 3000 ms, saturation offset = 3 ppm, see section 2.5), and the right column represents the same schedule, but with shorter saturation time and excitation flip angle (FA = 15°, and $T_{sat}$ = 1500 ms). The same CEST properties of scenario 'C' were used for all trajectories.

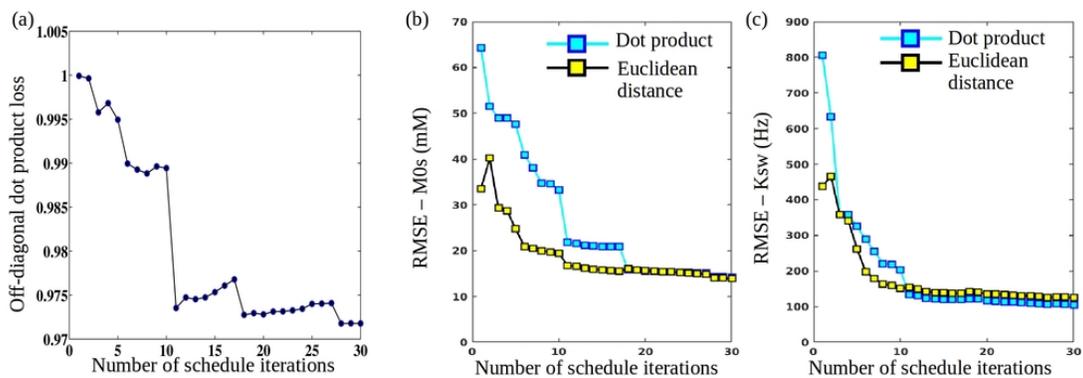

Figure 5: Optimization of the schedule length. (a) DP$_{loss}$ values for the baseline schedule with varied lengths. Note the step-like shape, predicting noticeable performance improvement at N$_t$ = 11, with further improvement when additional iterations are added. (b) Solute concentration (M0s) RMSE, comparing the dot-product and Euclidean-distance-based matching. (c) Proton exchange-rate (K$_{sw}$) RMSE for both matching metrics.

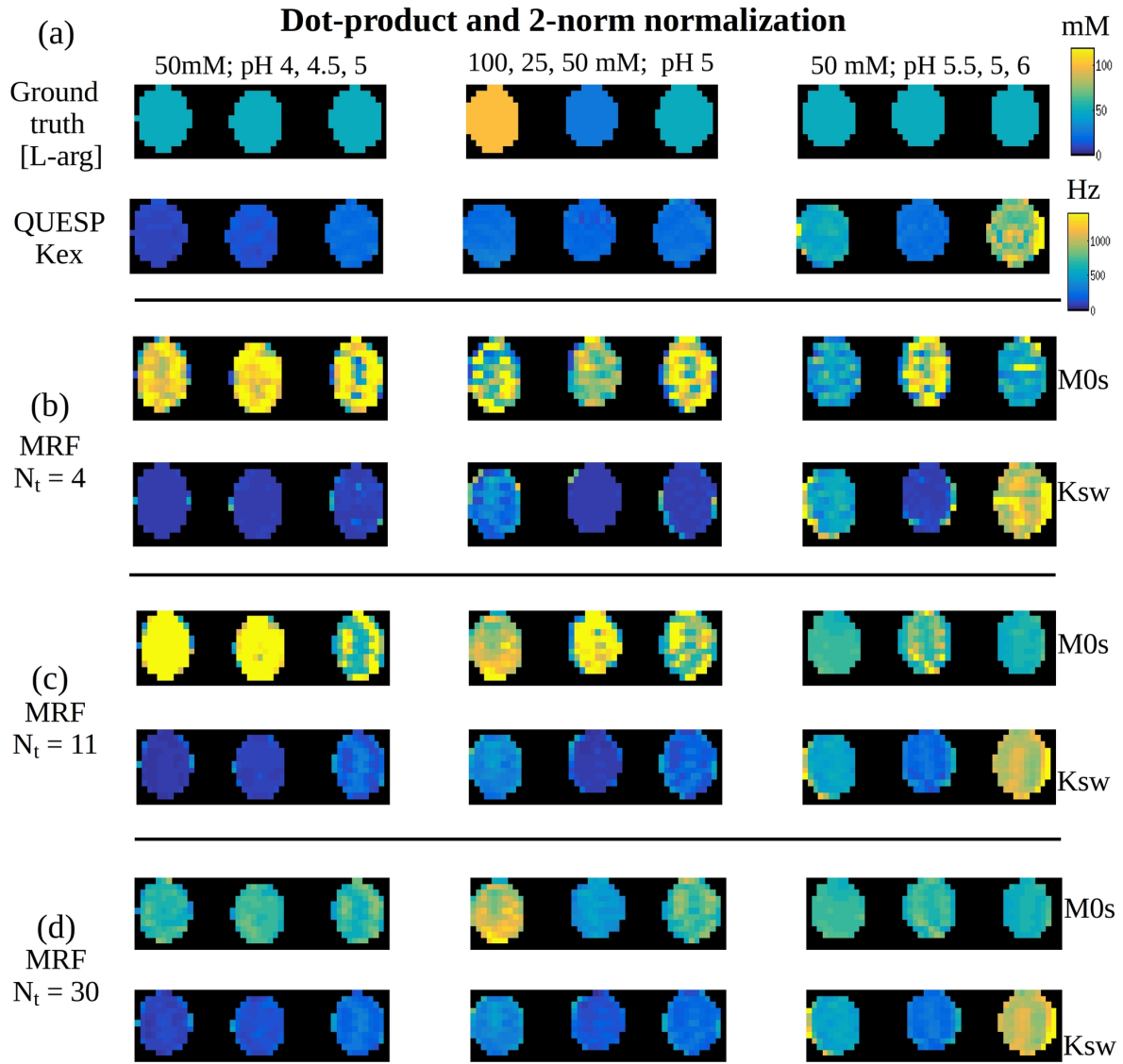

Figure 6: Dot-product matching of CEST-MRF phantom images. (a) Reference of known solute concentration (top) and QUESP proton exchange-rate images (bottom) for the 9 imaged vials. (b-d) Dot-product reconstructed CEST-MRF images for 4, 11, and 30 iterations respectively. The same color-map and dynamic range (top-right) were used for all images.

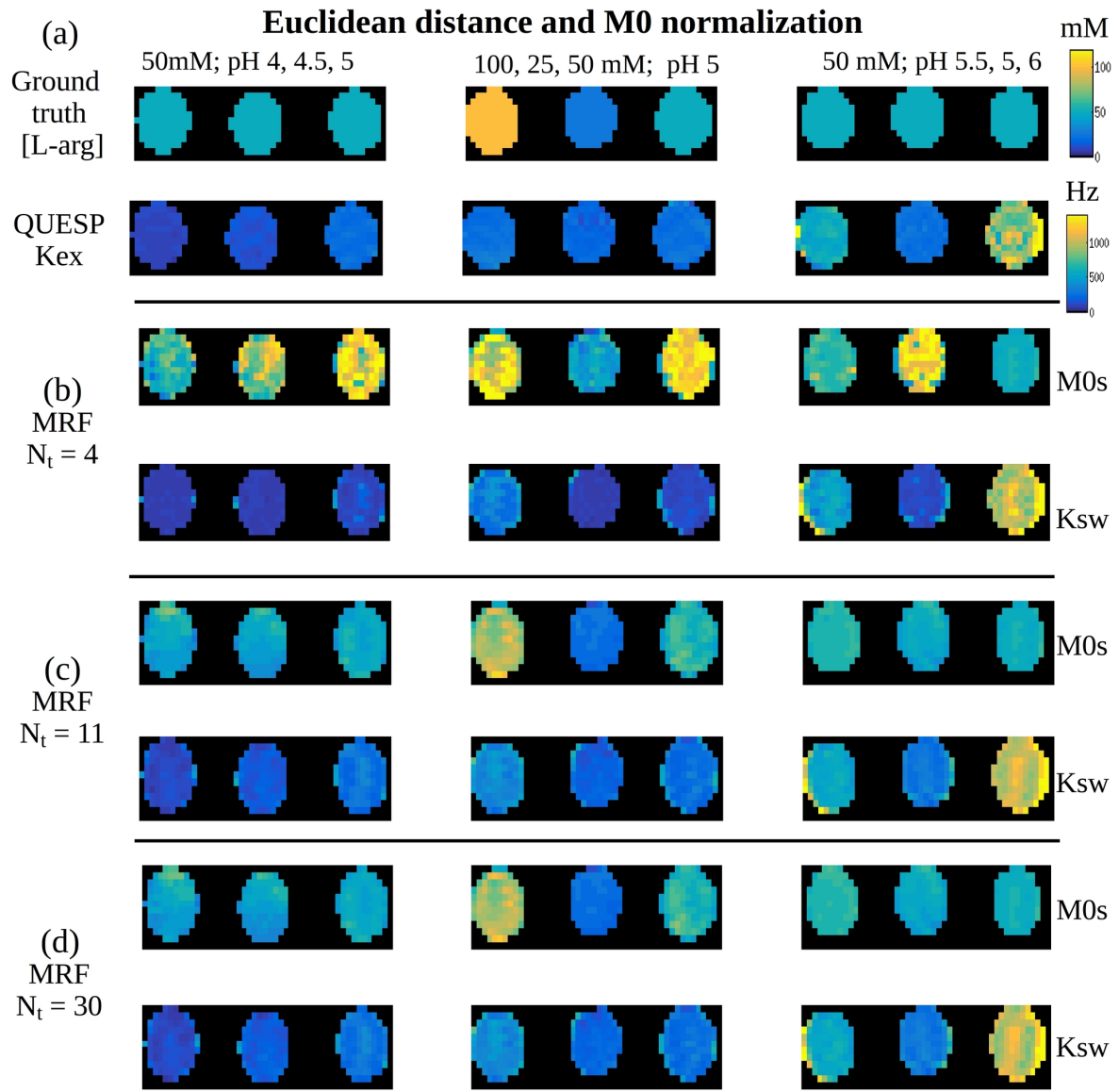

Figure 7: Euclidean-distance matching of CEST-MRF phantom images. (a) Reference of known solute concentration (top) and QUESP proton exchange-rate images (bottom) for the 9 imaged vials. (b-d) Euclidean-distance reconstructed CEST-MRF images for 4, 11, and 30 iterations respectively. The same color-map and dynamic range (top-right) were used for all images.

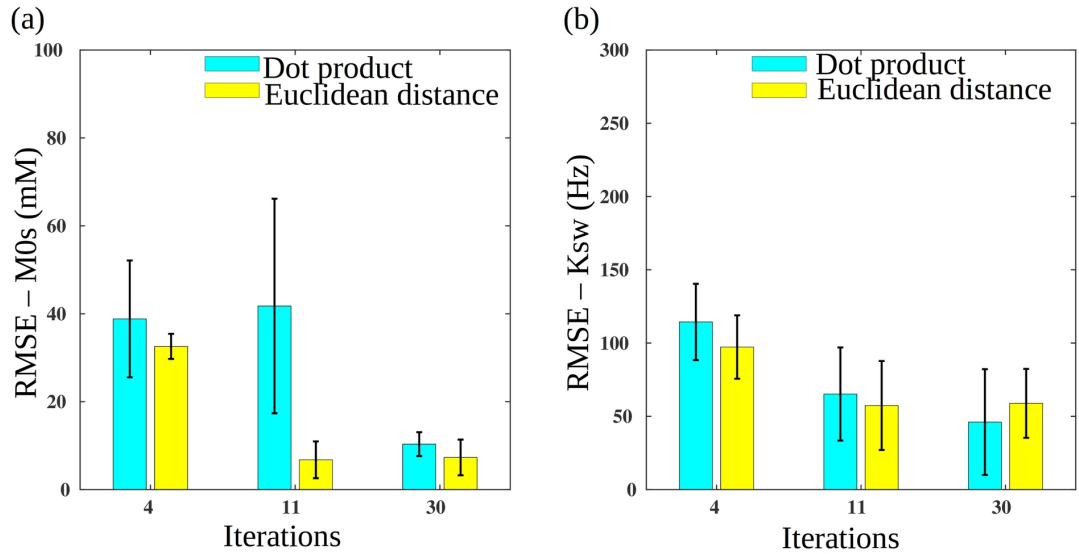

Figure 8: Phantom study quantitative analysis. RMSE values for solute concentration (M0s) (a), and chemical exchange-rate ($K_{sw}$) (b), using the baseline schedule with varied lengths. Note the significant performance improvement at $N_t = 11$ for Euclidean-distance M0s matching (a), significantly lower than the dot-product respective values. No significant improvement occurs for the Euclidean-distance at $N_t = 30$, suggesting the schedule can be shortened without harming performance.

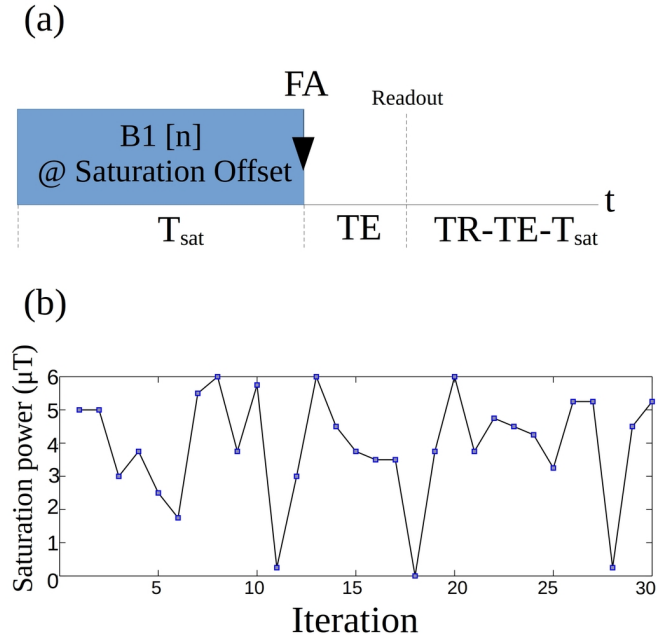

Figure S1: (a) Schematic of the CEST-MRF pulse sequence with the parameters optimized throughout this work. Each iteration comprises a continuous saturation block where a pseudo-random saturation power $B_1[n]$ is used at a specific saturation offset for a fixed saturation duration ($T_{sat}$). Next, standard imaging and relaxation blocks are performed, with a fixed flip angle (FA), TE, and TR. (b) The baseline saturation powers used in (42).